\documentclass[%
aps,prd,
amsmath,amssymb,amsfonts,
reprint,
superscriptaddress,
]{revtex4-2}

\usepackage{graphicx}
\usepackage{dcolumn}
\usepackage{bm}
\usepackage{arydshln}   
\usepackage{subcaption} 
\usepackage{soul}
\usepackage{color}
\usepackage{xcolor}
\usepackage{float}
\usepackage[colorlinks=true, allcolors=blue]{hyperref}
\usepackage{cleveref}
\usepackage{tikz}
\usepackage{enumitem}
\usepackage{multirow}
\usepackage{booktabs}
\usepackage{hyperref}

\begin{document}

\preprint{APS/123-QED}

\title{Investigating the $H_0$ Tension and Expansion-History Mismatch with Diverse Dark Energy Parametrization Frameworks}

\author{Upala Mukhopadhyay}
\email{upala.mukhopadhyay@uni.lu}
\affiliation{Department of Physics and Materials Science, University of Luxembourg, L-1511 Luxembourg City, Luxembourg}

\author{Purba Mukherjee}
\email{pdf.pmukherjee@jmi.ac.in}
\affiliation{Centre for Theoretical Physics, Jamia Millia Islamia, New Delhi - 110025}
\affiliation{Korea Astronomy and Space Science Institute, Daejeon 34055, Republic of Korea}

\author{Alexandre Tkatchenko}
\email{alexandre.tkatchenko@uni.lu}
\affiliation{Department of Physics and Materials Science, University of Luxembourg, L-1511 Luxembourg City, Luxembourg}



\begin{abstract}
In this work, we investigate whether the mismatch between early- and late-Universe inferences of the cosmic expansion history is associated with a particular redshift range, and whether models that alleviate the $H_0$ tension also provide a consistent description of $H(z)$ across cosmic time. We consider four late-time cosmological frameworks: $\Lambda$CDM, a dark energy (DE) equation-of-state parametrization, a DE pressure parametrization, and a generalized scale-factor parametrization. Using Planck CMB data as an early-Universe anchor and Pantheon+ with SH0ES calibration as a late-Universe probe, together with DESI DR2 BAO measurements, we first assess the $H_0$ tension in each framework and then reconstruct the expansion history through $H(z)$. We further compare these parametric reconstructions with non-parametric reconstruction employing Gaussian process regression.  We find that the expansion histories inferred within $\Lambda$CDM remain significantly discrepant over the full redshift range considered, whereas the alternative phenomenological parametrizations largely localize the mismatch to lower redshifts. Among these models, the pressure parametrization gives the strongest reduction of the $H_0$ tension, to approximately $2.6\sigma$, while still exhibiting significant discrepancies in the reconstructed $H(z)$ at intermediate redshifts. Both the parametric and model-independent reconstructions identify the strongest deviations around $z=0.51$ and $z=0.706$ consistently, indicating robust late-time
tensions in the cosmic expansion rate. 

\end{abstract}

\maketitle


\section{\label{introduction}Introduction}

The standard model of cosmology, the $\Lambda$CDM, which combines a cosmological constant $\Lambda$ \cite{Carroll:2000fy, Peebles:2002gy} with cold dark matter (CDM) within the framework of general relativity, has achieved remarkable success in describing a wide range of cosmological observations \cite{Huterer:2017buf,Bamba:2012cp}. Despite its empirical success, $\Lambda$CDM faces increasing challenges from the growing precision of modern cosmological data \cite{CosmoVerseNetwork:2025alb, Abdalla:2022yfr, Das:2013sca, DiValentino:2020vhf, DiValentino:2020zio, DiValentino:2020vvd, DiValentino:2020srs}. 
These include discrepancies among measurements from the Cosmic Microwave Background (CMB) \cite{Planck:2018vyg, ACT:2020gnv, Tristram:2023haj}, Type Ia supernovae \cite{Riess:2021jrx, Brout:2022vxf, Galbany:2022zir}, large-scale structure and galaxy surveys \cite{BOSS:2014hhw, BOSS:2016wmc, Addison:2017fdm, DESI:2024lzq, DESI:2024uvr}.

A particularly persistent discrepancy concerns the present-day value of the Hubble parameter, $H_0$, inferred from early- and late-Universe observations \cite{Riess:2021jrx}. Early-Universe constraints from Planck CMB observations \cite{Planck:2018vyg} provide a precise value of $H_0=67.4\pm0.5$ km s$^{-1}$ Mpc$^{-1}$ within flat $\Lambda$CDM cosmology, whereas late-time measurements from the SH0ES distance ladder \cite{Riess:2021jrx, Reid:2019tiq,HST:2000azd,Riess:2024vfa} yield $H_0=73.16\pm0.86$ km s$^{-1}$ Mpc$^{-1}$. 
The persistence of this discrepancy has motivated a wide range of theoretical proposals, including late dark energy \cite{Yang:2021flj,Copeland:2006wr,Najafi:2024qzm, Giare:2024gpk,Sahlu:2023wvl,Adil:2023ara,Menci:2024rbq,Bento:2002ps,Adil:2023exv,Vazquez:2023kyx,Vagnozzi:2019ezj,Du:2025csv,Yang:2025oax,Pedrotti:2025ccw,Mukherjee:2025myk}, early dark energy \cite{Kamionkowski:2022pkx, Poulin:2023lkg,McDonough:2023qcu,Niedermann:2019olb}, interacting dark matter and dark energy \cite{Wang:2016lxa,Farrar:2003uw, Amendola:1999er, Mukhopadhyay:2019jla, Mukhopadhyay:2020bml,Yang:2018euj,Wang:2024vmw,DiValentino:2019ffd,DiValentino:2019jae,Li:2024qso,Li:2026xaz,Dai:2026pvx,Shah:2024rme,Shah:2025ayl,Zhang:2025dwu}, and modified gravity theories \cite{Amendola:2006we,Nojiri:2022ski,Kobayashi:2019hrl,Sebastiani:2016ras,Mukhopadhyay:2019cai,Mukhopadhyay:2019wrw,Odintsov:2020qzd, Nojiri:2021dze, Yang:2024tkw, Silva:2023wvs, Langlois:2018dxi}. 
A comprehensive discussion of these proposals and their limitations can be found in Ref.~\cite{CosmoVerseNetwork:2025alb}.

Although this discrepancy is commonly characterized by the value of $H_0$ at the redshift $z=0$, it is not clear how the corresponding mismatch in the inferred expansion history evolves with redshift. Identifying the epoch at which deviations between early- and late-time inferences in $H(z)$ emerge is crucial for diagnosing the physical origin of the discrepancy. 
We therefore ask in this work, whether the expansion histories inferred from early- and late-Universe observations remain discrepant throughout cosmic evolution or whether the mismatch is concentrated within some particular redshift ranges.

To examine this possibility, we consider phenomenological frameworks that allow for deviations from the standard $\Lambda$CDM cosmology while remaining agnostic about the underlying microphysics. Our analysis focuses on the dark energy (DE) sector, which governs the late-time expansion of the Universe. Despite its dominant role, the physical nature of DE and its possible evolution with redshift remain unknown, making it a particularly compelling sector to explore possible modifications of the expansion history. Rather than committing to a specific theoretical realization of DE, we adopt general parametrizations of key physical quantities that characterize its evolution and, consequently, the expansion history of the Universe. In addition to the cosmological constant, we consider parametrizations of the (1) DE equation of state, (2) DE pressure density, and (3) scale factor of the Universe. 
These parametrizations are used as diagnostic frameworks to trace the inferred expansion history at different redshifts and to assess whether the additional late-time modifications change the distribution of the early--late mismatch across cosmic time.

We use Planck CMB measurements \cite{Planck:2018vyg} together with DESI DR2 baryon acoustic oscillation (BAO) data \cite{DESI:2025zgx} for the early-Universe inference, while the late-time inference is obtained from the Pantheon+ \cite{Scolnic:2021amr} supernova sample combined with the SH0ES \cite{Brout:2022vxf} local distance-ladder measurement and DESI DR2 BAO data. 
Within each framework, we compare the corresponding expansion histories reconstructed from the two dataset combinations and quantify their differences across redshift, including the present-day $z=0$. 
We further use a non-parametric reconstruction of $H(z)$ as a robustness test of whether the inferred redshift localization of the early-late mismatch depends on the assumed functional form.

The paper is organized as follows. Section~\ref{DE_model} introduces the phenomenological models, and Section~\ref{data} describes the observational datasets. Sections~\ref{result_DE1} presents the constraints of the cosmological parameters at the present-day $z=0$ and \ref{result_DE2} shows the reconstructed expansion histories across different redshifts. Section~\ref{result_DE3} compares the parametric and non-parametric reconstructions and finally, Section~\ref{Summary} summarizes our conclusions.

\section{\label{DE_model} {Models and Parameterizations}} 

We consider the following phenomenological models to trace the Hubble parameter $H(z)$ across different cosmic epochs. 
Rather than committing to a specific microphysical model, we adopt general parametrizations of quantities relevant to the cosmic expansion history:

\begin{enumerate}[left=0pt]

\item Cosmological Constant - Cold Dark Matter ($\Lambda$CDM): 

The $\Lambda$CDM model combines the cosmological constant $\Lambda$ with a constant equation of state (EoS) $w_\Lambda =-1$ \cite{Carroll:2000fy, Peebles:2002gy} and the cold dark matter component. 
The corresponding Hubble parameter is,
\begin{eqnarray}
 H^2(z) = H^2_0 \left[\Omega_{m0}(1+z)^3 + \Omega_{r0}(1+z)^4 + \Omega_{\Lambda0}\right]\,\,,
\end{eqnarray}
where $H_0$ is the Hubble parameter at the present epoch $z=0$. The present-day energy-density parameter of the component `$x$' is defined as $\Omega_{x0} = \rho_{x0}/\rho_{c0}$, where $\rho_{c0}$ ($=3H_0^2/(8\pi G)$) and $\rho_{x0}$ denote, respectively, the current critical energy density of the Universe and the current energy density of the component `$x$'. Here, $m$, $r$, and $\Lambda$ refer to the matter (baryon + dark matter), radiation, and cosmological-constant components, respectively. For a spatially flat Friedmann--Lemaitre--Robertson--Walker (FLRW) Universe, we have $\Omega_{\Lambda 0} = 1 - \Omega_{m0} - \Omega_{r0}$.

\item Equation-of-state parametrization of DE:

The Chevallier-Polarski-Linder (CPL) parametrization is a widely used phenomenological description of the DE equation of state (EoS), providing a two-parameter characterization of evolving DE \cite{Chevallier:2000qy, Linder:2002et}.


The CPL model is defined as 
\begin{equation}
    w_{\rm DE}(a) = w_0 + (1-a)\,w_a\,,
\end{equation}
where $a = (1+z)^{-1}$ denotes the scale factor, and $w_0$ and $w_a$ are constants, with $w_0$ representing the present-day DE EoS and $w_a$ characterizing its evolution. 
For a spatially flat Universe, the corresponding Hubble rate is
\begin{eqnarray}
    &H^2(z) = H_0^2 \left[
        \Omega_{m0}(1+z)^3
        + \Omega_{r0}(1+z)^4 + \right. ~~~~~\nonumber\\
          &~~~~~~~~~~+ \left. \Omega_{{\rm DE}0}(1+z)^{3(1+w_0+w_a)}
        \exp\!\left(-3w_a\,\frac{z}{1+z}\right)
    \right]\, ,
    \label{Hz_CPL}
\end{eqnarray}
where the DE energy density parameter $\Omega_{\mathrm{DE}0}$ satisfies $\Omega_{\mathrm{DE}0} = 1 - \Omega_{m0} - \Omega_{r0}$. The model reduces to $\Lambda$CDM for $w_0 = -1$ and $w_a = 0$.

\item Pressure-density parametrization of DE:

We consider a pressure parametrization (PP) framework of dynamical DE, following Refs. \cite{Sen:2007gk, Cheng:2025lod}. Instead of parameterizing the DE equation of state $w_{\rm DE}$, we use a Taylor expansion of the DE pressure density around the cosmological constant behaviour,
\begin{equation}
p_{\mathrm{DE}} \equiv p(a) = - p_0 + \sum\limits_{n \geq 1}\frac{(1 - a)^n}{n!}p_n\,\,, \label{PP taylor}
\end{equation}
where $n$ denotes the truncation order. Since $p=-p_0$ corresponds to the $\Lambda$CDM limit, we take $p_0 > 0$, with no assumption on the sign of $p_n$. 
Directly parameterizing the DE pressure is motivated by its explicit role in cosmic acceleration \cite{Sen:2007gk, Cheng:2025lod, Alam:2003sc}. A Lagrangian realization and a related normalized-density parametrization were discussed in Refs.~\cite{Sen:2007gk, Montefalcone:2026iga}.

We consider the first-order ($n=1$) expansion 
of Eq.~\ref{PP taylor}, for which the Hubble parameter becomes
\begin{eqnarray}
 &H^2(z) = H^2_0 \left[\Omega_{m 0}(1+z)^3 + \Omega_{r 0}(1+z)^4 + \right. \nonumber\\
 &\left.  + \Omega_{\rm DE 0} - \Omega_1\frac{z}{1+z}\right]\,\,. \label{HZ_pp1}  
\end{eqnarray}
Here, $\Omega_1 = \frac{3}{4}\frac{p_1}{3H^2_0 /8\pi G}$ and $\Omega_{\rm DE 0} = \frac{\rho_{\rm DE 0}}{3H^2_0 /8\pi G}$, where $\rho_{\rm DE 0}$ denotes the present-day DE energy density. This parametrization reduces to the standard $\Lambda$CDM when $p_1 = 0$. From the continuity equation ${\rm d}\rho_{\rm{DE}}/{\rm d}t + 3H(p_{\rm{DE}}+\rho_{\rm{DE}})=0$, the DE energy density and pressure density are $\rho_{\rm DE} = \rho_{\rm DE 0} - \frac{3}{4}(1- \frac{1}{1+z})p_1$ and $p_{\rm DE} = -\rho_{\rm DE 0} + \left(\frac{3}{4} -\frac{1}{1+z} \right) p_1$, respectively, giving the EoS of DE as,
\begin{equation}
    w_{\rm DE} = -1+\frac{1}{3}\frac{\Omega_1 \frac{1}{1+z}}{\Omega_1 \frac{z}{1+z} - \Omega_{\rm DE 0}}\,\,.\label{wde_pp1}
\end{equation}

\item Scale-factor parametrization:

The scale factor $a(t)$ provides the fundamental description of cosmic expansion in an FLRW Universe, with its time evolution determining the Hubble parameter through $H=\dot a/a$. Motivated by this, we consider the scale-factor parametrization proposed in Ref.~\cite{Sen:2001xu}. In the standard $\Lambda$CDM model, the exact analytical form of the scale factor at late times can be written as
$a(t) = a_0^{1/3} [\sinh (t/\tau)]^{2/3}$, where $\tau$ is a parameter with the dimension of time \cite{Reid:2002kp, Gron:2002wrd}. In Ref. \cite{Mukhopadhyay:2024fch, Choudhury:2025bnx}, the authors have added more flexibility to $a(t)$ in the following way
$a(t) = a_{1}^{2/\beta} [\sinh (t/\tau)]^{\beta}$, where $a_{1}$ and $\beta$ are dimensionless parameters. Including the contribution from radiation, the corresponding Hubble parameter is
\begin{eqnarray}
&H^{2} (z) =  H_{0}^{2} \left[A (1+z)^{2/\beta} + \Omega_{r0}(1+z)^4 + \Omega_{\rm{DE0}} \right]\,\,.\label{Hz_GM}
\end{eqnarray}
Here, $A = \frac{a_{1}^{2/\beta}}{1+a_{1}^{2/\beta}}$ and $\Omega_{\rm DE0} = 1-A-\Omega_{r0}$. 
Since Eq.~\ref{Hz_GM} describes deviations from $\Lambda$CDM without specifying whether they arise from DE, dark matter, or modified gravity, we refer to this as the General Model (GM), following Ref.~\cite{Mukhopadhyay:2024fch}.

The $\Lambda$CDM limit is recovered for $\beta= 2/3$ and $A=\Omega_{m0}$. Now, if the deviation from $\Lambda$CDM described in Eq. \ref{Hz_GM} appears due to DE behaviour, the energy density of DE will be
$\Omega_{\rm DE} = A (1+z)^{2/\beta} + (1-A-\Omega_{r0}) - \Omega_{m0}(1+z)^3$ and the corresponding EoS is \begin{eqnarray}
&w_{\rm DE}(z)=-1+
\frac{
\frac{2A}{\beta}(1+z)^{2/\beta}
-
3\Omega_{m0}(1+z)^3
}{
3\left[
A(1+z)^{2/\beta}
+
\left(1-A-\Omega_{r0}\right)
-
\Omega_{m0}(1+z)^3
\right]
} \, . \, \, \label{EOS_GM}
\end{eqnarray}

\end{enumerate}

\section{\label{data} Observational Data Sets}

We use recent observations of Type Ia supernovae (SNe Ia), baryon acoustic oscillations (BAO), and the cosmic microwave background (CMB) to constrain the model parameters. 
Together, these probes constrain the expansion history over complementary redshift ranges.

\begin{enumerate}[left=0pt]

\item {\bf Pantheon+SH0ES:} Distance modulus measurements of Type Ia supernovae from the Pantheon+ sample \cite{Brout:2022vxf} include 1701 light curves from 1550 distinct SNe Ia spanning $z\in[0.001,2.26]$. We also adopt the SH0ES Cepheid calibration and its associated covariance matrix. We refer to this combined dataset as Pantheon+SH0ES.

The distance modulus $\mu(z)$ is related to the luminosity distance $d_L(z)$ through
$\mu(z) = m_b(z) - M_b = 5 \log_{10} \left( \frac{d_L(z)}{\mathrm{Mpc}} \right) + 25 ,$
where $m_b$ and $M_b$ are the apparent and absolute magnitudes, respectively. In a spatially flat FLRW Universe,
$d_L(z)=(1+z)\int_0^z \frac{c}{H(z')}\,{\rm d}z' .$
We treat $M_b$ as a nuisance parameter calibrated using the SH0ES Cepheid distance ladder.

\item {\bf DESI:} We use BAO measurements from DESI DR2 \cite{DESI:2025zgx}. The DESI survey provides BAO signal in both the radial and transverse directions. The relevant anisotropic distance measures are $D_M(z)=\int_0^z \frac{c}{H(z')}\,{\rm d}z',$ and $D_H(z)=\frac{c}{H(z)}.$ DESI also provides the isotropic volume-averaged distance $D_V(z)=\left[z\,D_M^2(z)\,D_H(z)\right]^{1/3}.$ All BAO measurements are expressed relative to the sound horizon at the drag epoch, $r_d$. We therefore use $D_M(z)/r_d$, $D_H(z)/r_d$, and $D_V(z)/r_d$. We refer to this dataset as DESI.

\item \textbf{Planck:} We use the Planck 2018 CMB temperature and polarization measurements \cite{Planck:2018vyg} which includes the high-$\ell$ TT, TE, and EE spectra, the low-$\ell$ TT and EE likelihoods, and the CMB lensing likelihood. We refer to this combination as Planck.
\end{enumerate}

We constrain the model parameters using Markov Chain Monte Carlo (MCMC) analyses with uniform priors over physically motivated ranges. For the low-$z$ combination, Pantheon+SH0ES+DESI, we use the \texttt{emcee} sampler \cite{Foreman-Mackey:2012any}, 
which implements the affine-invariant ensemble MCMC algorithm. For the high-$z$ analysis involving CMB data, we use the Boltzmann solver \texttt{CLASS} \cite{Blas:2011rf} together with \texttt{MontePython} \cite{Brinckmann:2018cvx}. 
Convergence is assessed using the Gelman--Rubin diagnostic \cite{Gelman:1992zz}, requiring $R-1<0.02$. Our analysis builds upon the standard six-parameter $\Lambda$CDM framework implemented in \texttt{CLASS}. These parameters are $\Omega_bh^2$, $\Omega_ch^2$, $\theta_s$, $\tau$, $\ln(10^{10}A_s)$, and $n_s$. The extended models considered here introduce additional parameters beyond the baseline ones: the CPL model includes $w_0$ and $w_a$, the PP model includes $\Omega_1$, and the GM model includes $A$ and $\beta$. The resulting MCMC chains are analyzed and visualized using \texttt{getdist} \cite{Lewis:2019xzd}, from which we obtain marginalized constraints and posterior distributions.

\begin{table*}[ht]
\centering
\renewcommand{\arraystretch}{1.2}
\setlength{\tabcolsep}{7pt}
\begin{tabular}{l l c c c}
\toprule
{\bf Model} & {\bf Parameter} & {\bf Planck+DESI} & {\bf Pantheon+SH0ES+DESI} & {\bf Tension} \\
\midrule
\multirow{5}{*}{$\Lambda$CDM}
 & $H_0$ & $67.58 \pm 0.24$ & $73.6 \pm 1.0$ & $5.85\sigma$ \\
 & $\Omega_{m0}$ & $0.3110 \pm 0.0032$ & $0.318 \pm 0.014$ & $0.49\sigma$ \\
 & $h\,r_d$ & $99.40 \pm 0.24$ & $101.350 \pm 2.066$ & $0.94\sigma$ \\
\midrule
\multirow{7}{*}{CPL}
 & $H_0$ & $63.7^{+1.8}_{-2.1}$ & $73.3 \pm 1.0$ & $4.38\sigma$ \\
 & $\Omega_{m0}$ & $0.351 \pm 0.022$ & $0.261^{+0.056}_{-0.021}$ & $2.03\sigma$ \\
 & $w_0$ & $-0.44 \pm 0.21$ & $-0.860 \pm 0.055$ &  \\
 & $w_a$ & $-1.68 \pm 0.59$ & $0.13^{+0.58}_{-0.36}$ &  \\
 & $h\,r_d$ & $93.938 \pm 2.953$ & $99.688 \pm 2.054$ & $1.60\sigma$ \\
\midrule
\multirow{6}{*}{PP}
 & $H_0$ & $69.3 \pm 1.1$ & $73.2 \pm 1.0$ & $2.62\sigma$ \\
 & $\Omega_{m0}$ & $0.2939 \pm 0.0090$ & $0.2937 \pm 0.0092$ & $0.02\sigma$ \\
 & $\Omega_1$ & $0.11^{+0.12}_{-0.11}$ & $-0.29 \pm 0.11$ &  \\
 & $h\,r_d$ & $102.370 \pm 1.631$ & $99.625 \pm 1.999$ & $1.06\sigma$ \\
\midrule
\multirow{6}{*}{GM}
 & $H_0$ & $68.97 \pm 0.63$ & $73.3 \pm 1.0$ & $3.66\sigma$ \\
 & $A$ & $0.292 \pm 0.013$ & $0.384^{+0.031}_{-0.036}$ &  \\
 & $\beta$ & $0.6651 \pm 0.0022$ & $0.722 \pm 0.021$ &  \\
 & $h\,r_d$ & $101.669 \pm 0.945$ & $93.799 \pm 1.987$ & $3.58\sigma$ \\
\bottomrule
\end{tabular}
\caption{Marginalized constraints at the 68$\%$ C.L. for the model and cosmological parameters, obtained from the combined datasets Planck+DESI and Pantheon+SH0ES+DESI. The corresponding tensions in $H_0$, $\Omega_{m0}$, and $h r_d$ for each model are also shown (note, $H_0$ is in units of km Mpc$^{-1}$ s$^{-1}$).}
\label{table_CL}
\end{table*}

\begin{figure*}
\centering
\includegraphics[width=0.48\textwidth]{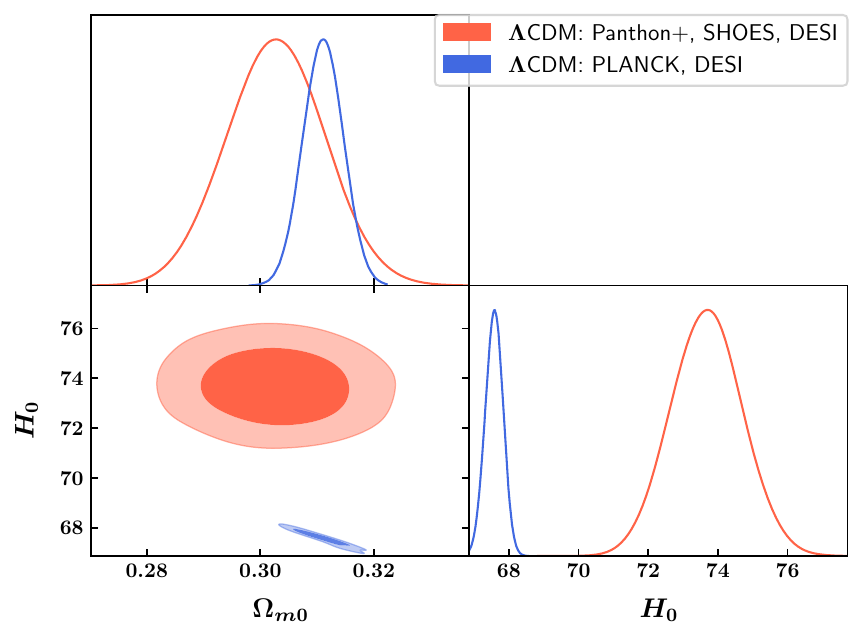}
\includegraphics[width=0.48\textwidth]{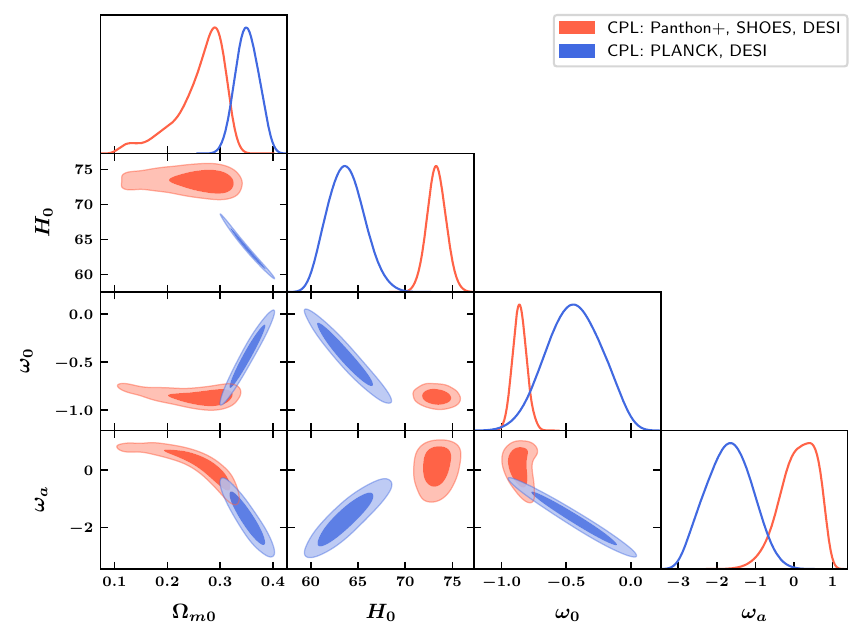}
\includegraphics[width=0.48\textwidth]{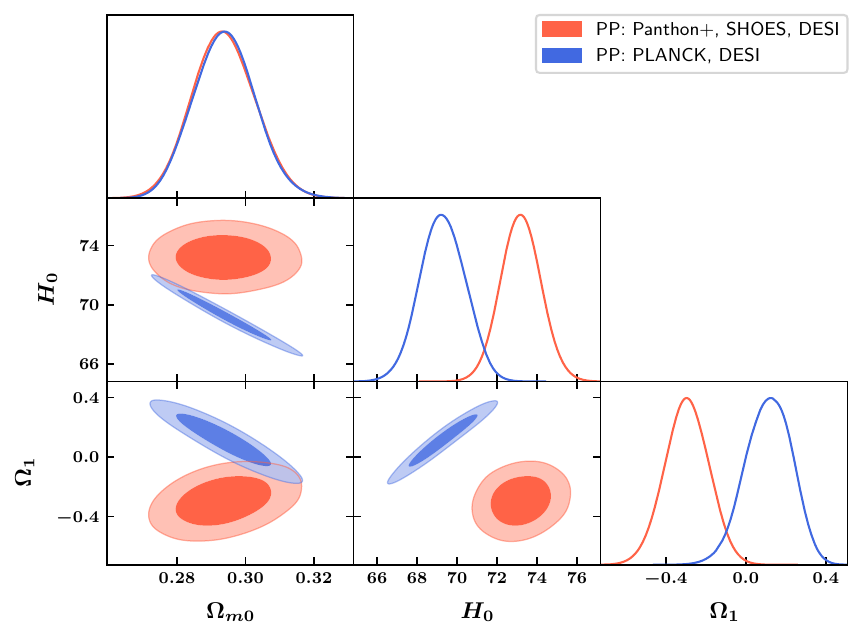}
\includegraphics[width=0.48\textwidth]{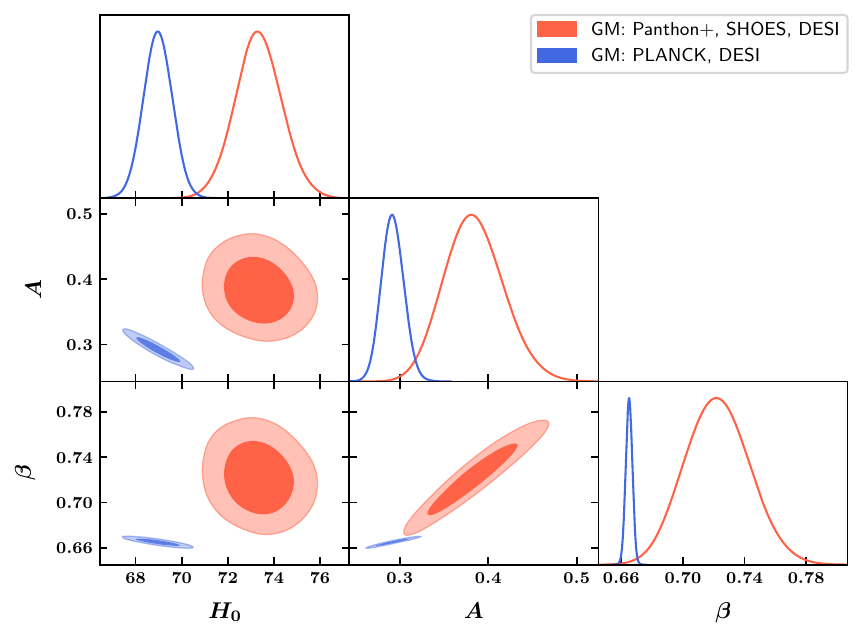}
\caption{One-dimensional posterior distributions and two-dimensional marginalized contours for (a) (upper left panel) $\Lambda$CDM, (b) (upper right panel) CPL model, (c) (lower left panel) PP model and (d) (lower right panel) GM model.}
\label{f:posterior distribution}
\end{figure*}

\begin{table*}[ht]
\centering
\renewcommand{\arraystretch}{1.15}
\setlength{\tabcolsep}{10pt}
\begin{tabular}{l c c c c}
\toprule
Model & \multicolumn{2}{c}{Planck+DESI} & \multicolumn{2}{c}{Pantheon+SH0ES+DESI} \\
\cmidrule(lr){2-3} \cmidrule(lr){4-5}
 & $\Delta\chi^2_{\rm min}$ & $\Delta$AIC & $\Delta\chi^2_{\rm min}$ & $\Delta$AIC \\
\midrule
CPL          & $-3.99$ & $0.01$ & $-2.70$  & $1.30$ \\
PP           & $-0.04$  & $1.96$  & $-2.50$  & $-0.50$ \\
GM           & $-3.10$ & $-1.1$ & $-2.80$ & $-0.80$ \\
\bottomrule
\end{tabular}
\caption{Relative goodness-of-fit statistics with respect to $\Lambda$CDM for the two dataset combinations. Negative values of $\Delta\chi^2_{\rm min}$ and $\Delta$AIC indicate a better fit than $\Lambda$CDM.}
\label{table_stats}
\end{table*}

\section{\label{result_DE1} Cosmic expansion rate at the present epoch}

In this section, we use the datasets introduced in Sect.~\ref{data} to constrain the phenomenological frameworks described in Sect.~\ref{DE_model}. 
For each model, we compare constraints obtained from the late-time combination Pantheon+SH0ES+DESI with those from early-time combination Planck+DESI. We first examine the present-day parameters, particularly $H_0$ and $\Omega_{m0}$, as a diagnostic of consistency between the two inferred expansion histories at $z=0$.
Table~\ref{table_CL} summarizes the marginalized constraints at the 68$\%$ C.L., while the corresponding posterior distributions and confidence contours are shown in Fig.~\ref{f:posterior distribution}. 
The main features of the parameter constraints are summarized below.

\begin{enumerate}
\item {\bf $\Lambda$CDM:}
For $\Lambda$CDM, shown in Fig.~\ref{f:posterior distribution}(a) ( upper left panel of Fig.~\ref{f:posterior distribution}), the low- and high-redshift constraints show no overlap in the $H_0$--$\Omega_{m0}$ plane at the $2\sigma$ level. The inferred values of $H_0$ differ by more than $5\sigma$, demonstrating a strong inconsistency between the present-day expansion rates inferred from the early- and late- dataset combinations.

\item {\bf CPL:}
The CPL constraints are shown in Fig.~\ref{f:posterior distribution}(b) (upper right panel of Fig.~\ref{f:posterior distribution}). Compared with $\Lambda$CDM, the discrepancy in $H_0$ is reduced to about $4.3\sigma$. 
This reduction is largely associated with broader parameter uncertainties, while the mean values inferred from the two datasets remain well separated. The inferred values of $\Omega_{m0}$ differ by more than $2\sigma$, and the corresponding $H_0$--$\Omega_{m0}$ contours remain disjoint at the $2\sigma$ level.

\item {\bf PP:}
The results for the PP model are shown in the lower left panel of Fig.~\ref{f:posterior distribution} (Fig.~\ref{f:posterior distribution}(c)).
In this case, the discrepancy in $H_0$ is reduced to about $2.6\sigma$. 
Unlike the CPL case, this reduction is primarily associated with a shift in the Planck+DESI mean value toward the late-time inference rather than with substantially enlarged uncertainties.
The two determinations of $\Omega_{m0}$ are also mutually consistent, with the corresponding $H_0$--$\Omega_{m0}$ contours approaching overlap at the boundary of the $2\sigma$ region.
This indicates that, among the models considered, the PP parametrization provides a promising possibility for addressing the $H_0$ tension while preserving consistency in $\Omega_{m0}$.

\item {\bf GM:}
The GM constraints are shown in Fig.~\ref{f:posterior distribution}(d) (lower right panel of Fig.~\ref{f:posterior distribution}). The discrepancy in $H_0$ is reduced relative to $\Lambda$CDM but remains at about $3.6\sigma$. We also find a shift in the parameter $\beta$ between the two dataset combinations. Planck+DESI gives
$\beta = 0.6651 \pm 0.0022$, consistent with the $\Lambda$CDM limit $\beta=2/3$, whereas Pantheon+SH0ES+DESI gives
$\beta = 0.722 \pm 0.021$, corresponding to a deviation of approximately $2.6\sigma$ from this limit. The two determinations of $\beta$ differ by about $2.7\sigma$. 
This indicates that the preference for a departure from the $\Lambda$CDM limit is driven by the late-time data combination.
\end{enumerate}

Table~\ref{table_CL} summarizes the marginalized constraints at the 68$\%$ C.L. for the model and cosmological parameters, obtained from the combined datasets Planck+DESI and Pantheon+SH0ES+DESI. Table~\ref{table_CL} also reports the corresponding tensions in $H_0$, $\Omega_{m0}$, and $h r_d$ for each model. 

We further compare the overall statistical performance of the models in Table~\ref{table_stats} using the minimum chi-square and the Akaike Information Criterion (AIC) \cite{Akaike:1974vps},
\begin{equation}
{\rm AIC}=\chi^2+2K,
\end{equation}
where $K$ denotes the number of free parameters. We report
$\Delta\chi^2 \equiv \chi^2_{\rm model}-\chi^2_{\Lambda{\rm CDM}}$
and
$\Delta{\rm AIC}\equiv{\rm AIC}_{\rm model}-{\rm AIC}_{\Lambda{\rm CDM}}$,
with negative values indicating an improved fit relative to $\Lambda$CDM. 
Overall, Table~\ref{table_stats} shows that the extended parametrizations provide fits statistically comparable to $\Lambda$CDM for the datasets considered. 
Thus, although the inferred present-day parameters differ appreciably across the models, the current data do not establish a decisive statistical preference for any extended framework over $\Lambda$CDM. The more important question is therefore whether these differences at $z=0$ persist, weaken, or become localized within some particular redshift ranges when the full expansion history $H(z)$ is considered. To further compare the phenomenological models described in Sect. \ref{DE_model}, we examine the evolutions of DE within each framework in Appendix \ref{appendix A}. 

\begin{figure*}
\centering
\includegraphics[width=0.48\textwidth]{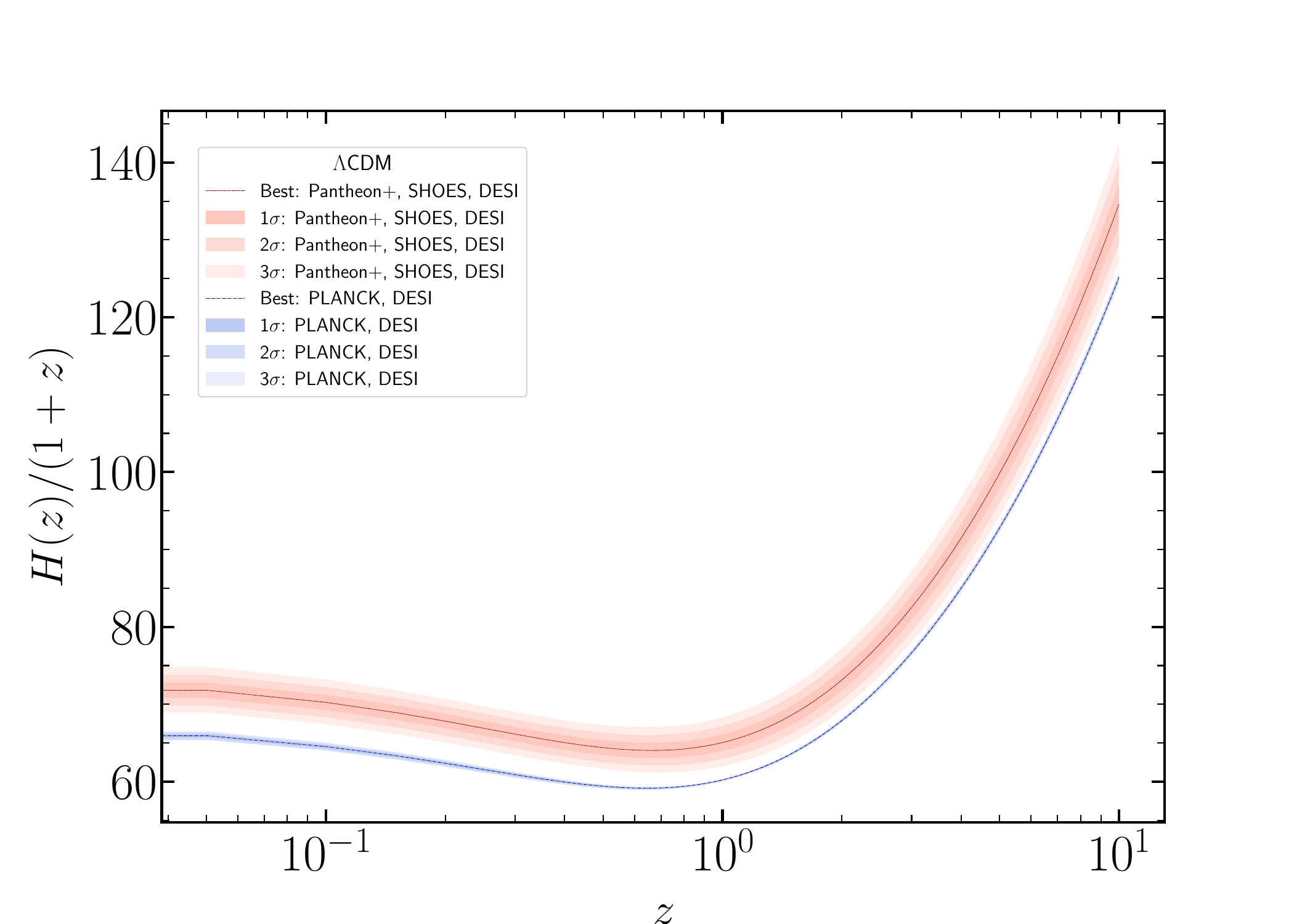}
\includegraphics[width=0.48\textwidth]{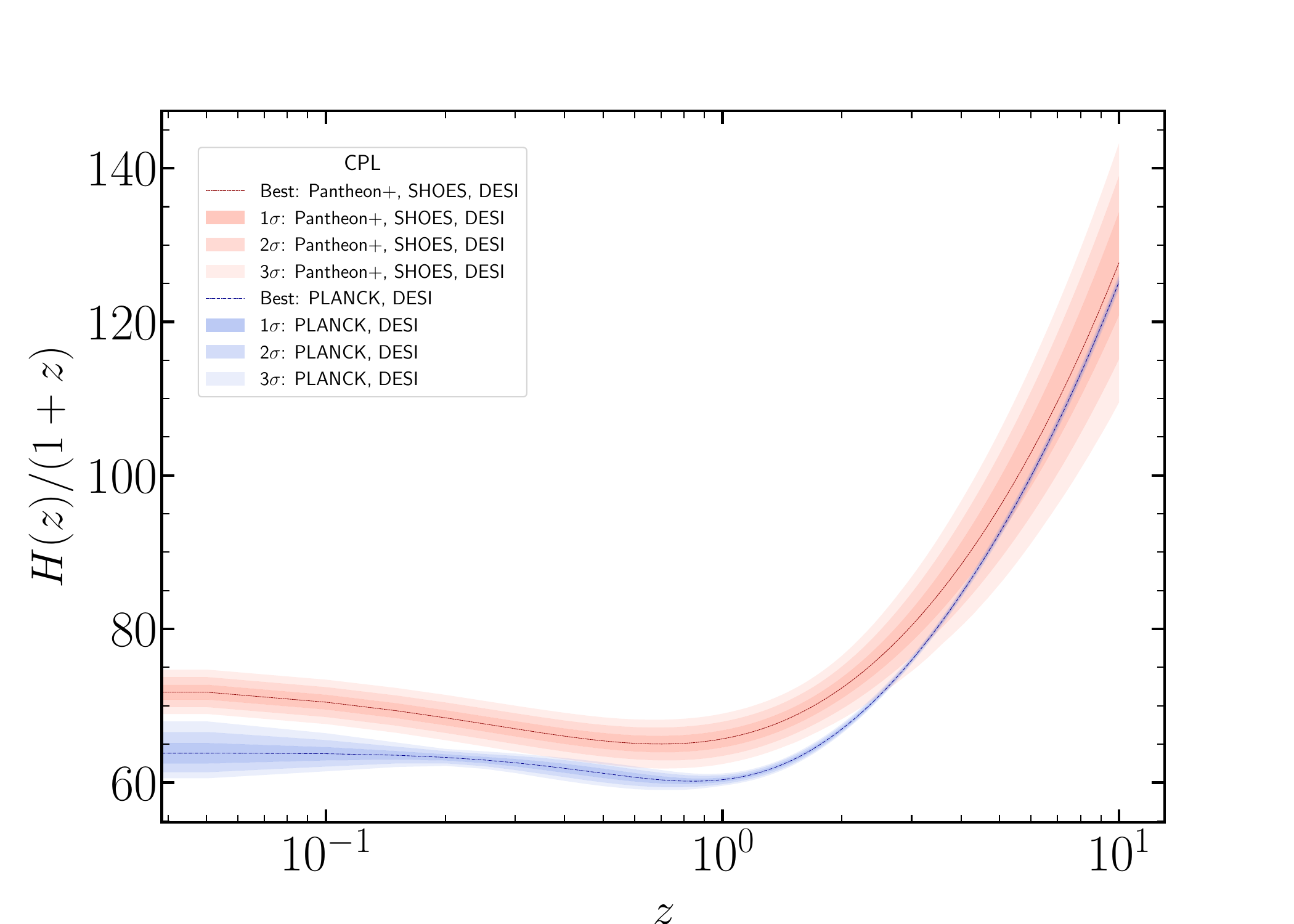}
\includegraphics[width=0.48\textwidth]{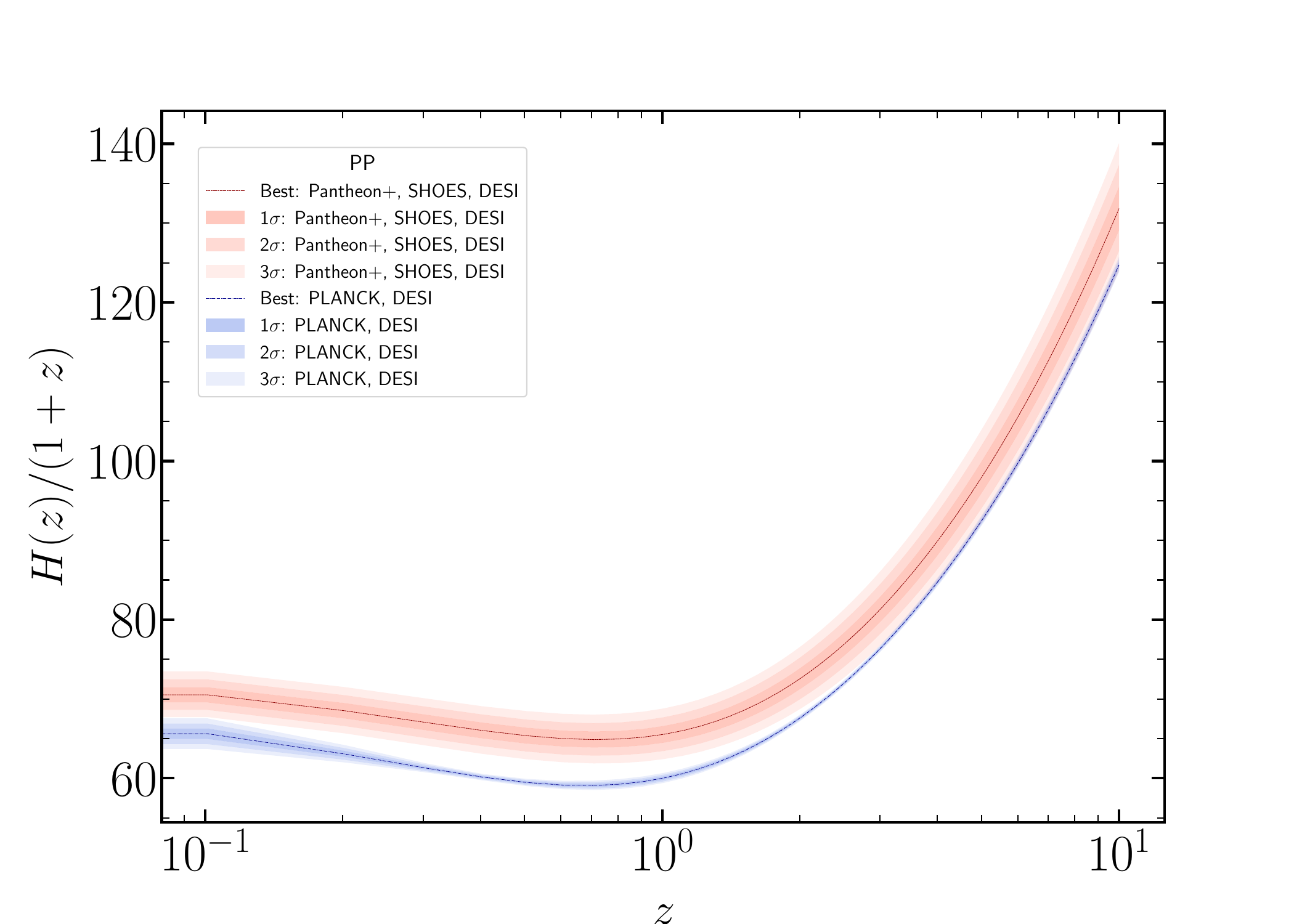}
\includegraphics[width=0.48\textwidth]{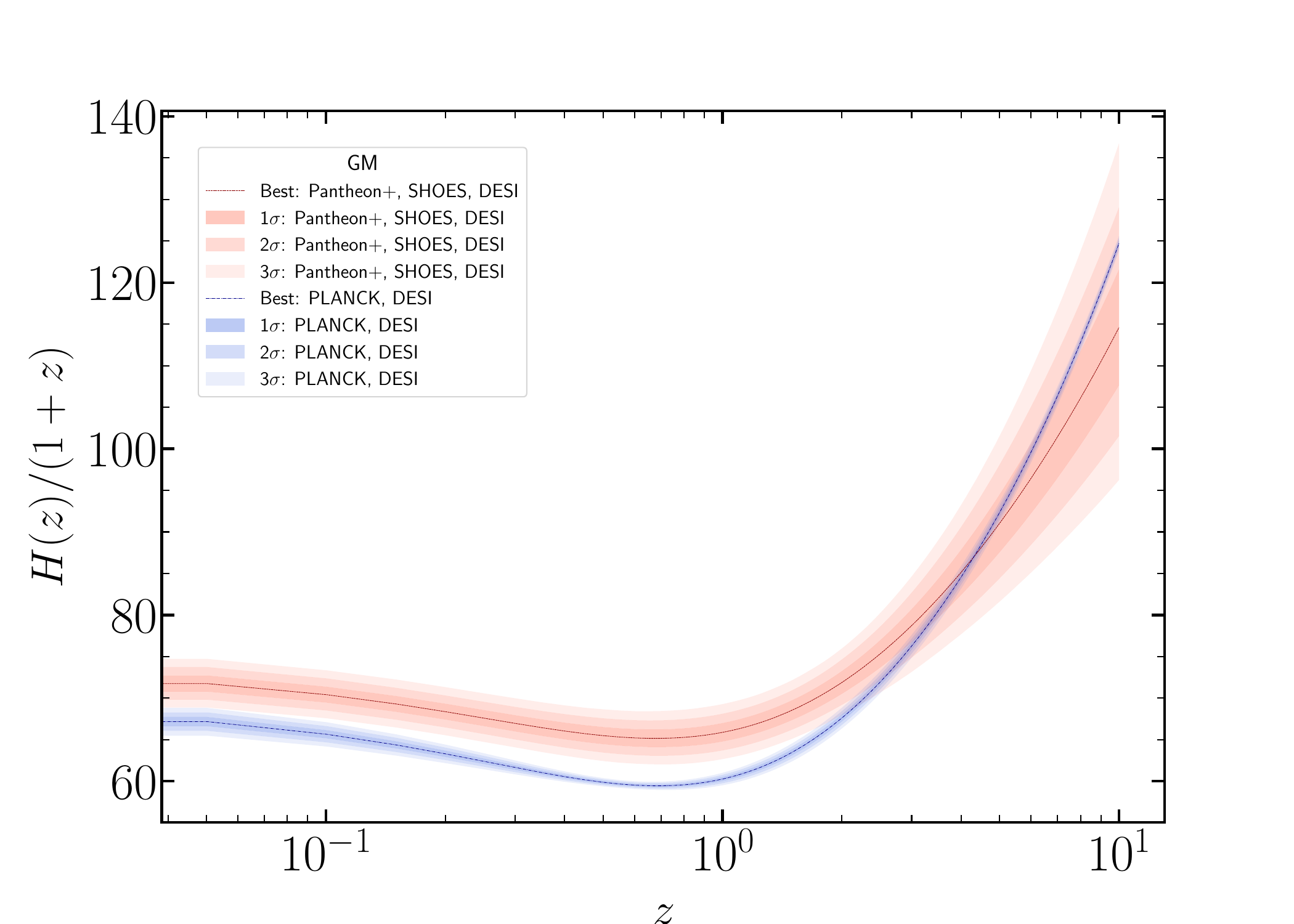}
\caption{Reconstructed $H(z)/(1+z)$ vs $z$ for (a) (upper left) $\Lambda$CDM, (b) (upper right) CPL, (c) (lower left) PP, (d) (lower right) GM for the redshift range $0 \leq z \leq 10$. The tomato and blue regions correspond to the allowed spaces for Pantheon+SH0ES+DESI and Planck+DESI, respectively. Here, H(z) is in unit of km Mpc$^{-1}$ s$^{-1}$.}
\label{f:plot_z}
\end{figure*}

\begin{figure}
  \centering
  \begin{subfigure}{0.5\textwidth}
      \centering
      \includegraphics[width=1.0\textwidth]{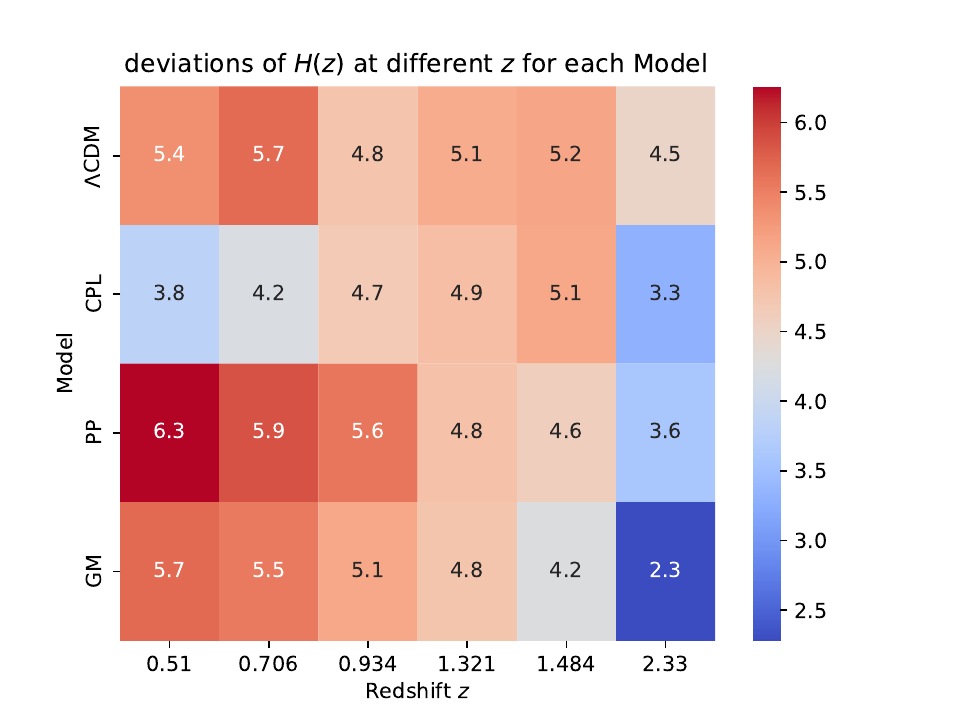}
        \end{subfigure}
\caption{Differences between $H(z)$ reconstructed from Pantheon+SH0ES+DESI and Planck+DESI at the DESI tracer redshifts for the $\Lambda$CDM, CPL, PP, and GM parametrizations. The colour scale shows the discrepancy in units of $\sigma$.}  
  \label{f:heatmap_Hz}
\end{figure}

\section{\label{result_DE2} Cosmic Expansion Rate through Cosmic Time}

In this section, we aim to study the 
discrepancy between the early- and late-time inferences of the Hubble parameter across different redshifts. We use the parameter constraints obtained in Sect.~\ref{result_DE1} to reconstruct the corresponding expansion histories $H(z)$ and investigate whether the mismatch is distributed broadly over cosmic time or concentrated within a particular redshift regime.

We first examine the continuous redshift evolution of the reconstructed expansion histories up to $z=10$.
The corresponding results are presented in Fig.~\ref{f:plot_z}. We show the reconstructed expansion rate $H(z)/(1+z)$, which is proportional to $\dot{a}$, across redshift for each model using the corresponding parameter constraints shown in Fig.~\ref{f:posterior distribution}. The minimum of $H(z)/(1+z)$ corresponds to the redshift at which the Universe transits from a decelerated to an accelerated phase of cosmic expansion.

\begin{enumerate}[left=0pt]
\item {\bf $\Lambda$CDM:} The reconstructed expansion history for $\Lambda$CDM is shown in Fig.~\ref{f:plot_z}(a) (upper left panel of Fig.~\ref{f:plot_z}). 
The tomato- and blue-coloured regions correspond to the Pantheon+SH0ES+DESI and Planck+DESI reconstructions, respectively. The two reconstructed expansion histories differ by more than $4\sigma$ throughout the plotted redshift range. 
The discrepancy also remains above $4\sigma$ when the reconstruction is extended to $z\sim1100$, indicating a broad mismatch across all redshifts.

\item {\bf CPL:} The reconstructed evolution for CPL is shown in Fig.~\ref{f:plot_z}(b) (upper right panel of Fig.~\ref{f:plot_z}). Unlike in $\Lambda$CDM, the differences between the two expansion histories are concentrated mainly at lower redshift. The reconstructions differ by about $3\sigma$ for $z\lesssim2.5$, with the separation becoming more pronounced toward lower redshift, while at higher redshift they overlap within $1\sigma$. 
Thus, relative to $\Lambda$CDM, the additional freedom in the CPL parametrization changes how the mismatch is distributed across redshift: the discrepancy is substantially reduced at higher redshift but remains significant at lower redshift.

\item {\bf PP:} The reconstructed PP expansion history is shown in Fig.~\ref{f:plot_z}(c) (lower left panel of Fig.~\ref{f:plot_z}). The two reconstructions remain within $3\sigma$ at higher redshift, while the discrepancy becomes more pronounced for $z\lesssim2$, grows toward lower redshift, and then weakens again for $z\lesssim0.1$. 
PP therefore improves the high-redshift agreement relative to $\Lambda$CDM, although less strongly than CPL.
We also note that extending the PP model to second order around the cosmological constant leads to qualitatively similar results. 

\item {\bf GM:} The reconstructed GM expansion history is shown in Fig.~\ref{f:plot_z}(d) (lower right panel of Fig.~\ref{f:plot_z}). The differences are most pronounced at lower redshift, remaining within $3\sigma$ for $z>2$ but exceeding $3\sigma$ for $z\lesssim2$. 
A comparison with the other parametrizations further reveals a qualitatively distinct behaviour for GM: 
for $z\gtrsim4$, the $H(z)$ values inferred from Planck+DESI are larger than those from Pantheon+SH0ES+DESI, whereas the ordering is reversed at lower redshift. 
This indicates that although DE is assumed to be a dominating component at low redshift, the assumed parametrization can still produce nontrivial differences in the reconstructed expansion history at higher redshifts.

\end{enumerate}

Since from Fig.~\ref{f:plot_z} in general we get that the differences between the expansion rates inferred from early- and late-time data larger when $z$ is less than 2.5, we next quantify the differences at the specific redshifts directly probed by the DESI tracers.

Figure~\ref{f:heatmap_Hz} illustrates the differences between the reconstructed $H(z)$ values for each model obtained from Pantheon+SH0ES+DESI and Planck+DESI at the DESI tracer redshifts $z=0.51$, $0.706$, $0.934$, $1.321$, $1.484$, and $2.33$. The colour code in Fig.~\ref{f:heatmap_Hz} represents the differences in units of $\sigma$ (which are computed as the difference between the mean values divided by the quadrature sum of the corresponding $1\sigma$ errors).

For the PP model, the discrepancies between the two reconstructions are $6.3\sigma$, $5.9\sigma$, $5.6\sigma$, $4.8\sigma$, $4.6\sigma$, and $3.6\sigma$ at these respective redshifts. Thus, despite reducing the present-day $H_0$ discrepancy to $\sim2.62\sigma$ (Table~\ref{table_CL}), the PP model exhibits substantial discrepancies exceeding $5\sigma$ at $z=0.51$, $0.706$, and $0.934$. This highlights that improved agreement in $H_0$ at $z=0$ does not necessarily imply consistency across the full expansion history.

Figure~\ref{f:heatmap_Hz} also shows that $\Lambda$CDM exhibits discrepancies larger than $4\sigma$ at all DESI tracer redshifts. In contrast, the remaining three models display a noticeably smaller mismatch toward higher redshift. In particular, at $z=2.33$, the deviations for $\Lambda$CDM, CPL, PP, and GM are $4.5\sigma$, $3.3\sigma$, $3.6\sigma$, and $2.3\sigma$, respectively.

In summary, the reconstructed $H(z)$ histories in Figs.~\ref{f:plot_z}(a)--\ref{f:plot_z}(d), together with the DESI-redshift comparison in Fig.~\ref{f:heatmap_Hz}, show that 
the redshift dependence of the early--late mismatch differs substantially across the parametrizations. In $\Lambda$CDM, the discrepancy remains broad in all redshifts, whereas the extended parametrizations concentrate the largest differences mainly at low redshifts. Therefore, the continuous reconstruction and the DESI tracer comparison together identify the low-redshift regime as the region where the discrepancies are most pronounced. The persistence of such anomalies at low-redshifts across multiple independent phenomenological models may indicate either the presence of new late-time physics or unaccounted-for systematic effects in low-redshift observations.

\begin{figure}
  \centering
  \begin{subfigure}{0.5\textwidth}
      \centering
      \includegraphics[width=1.0\textwidth]{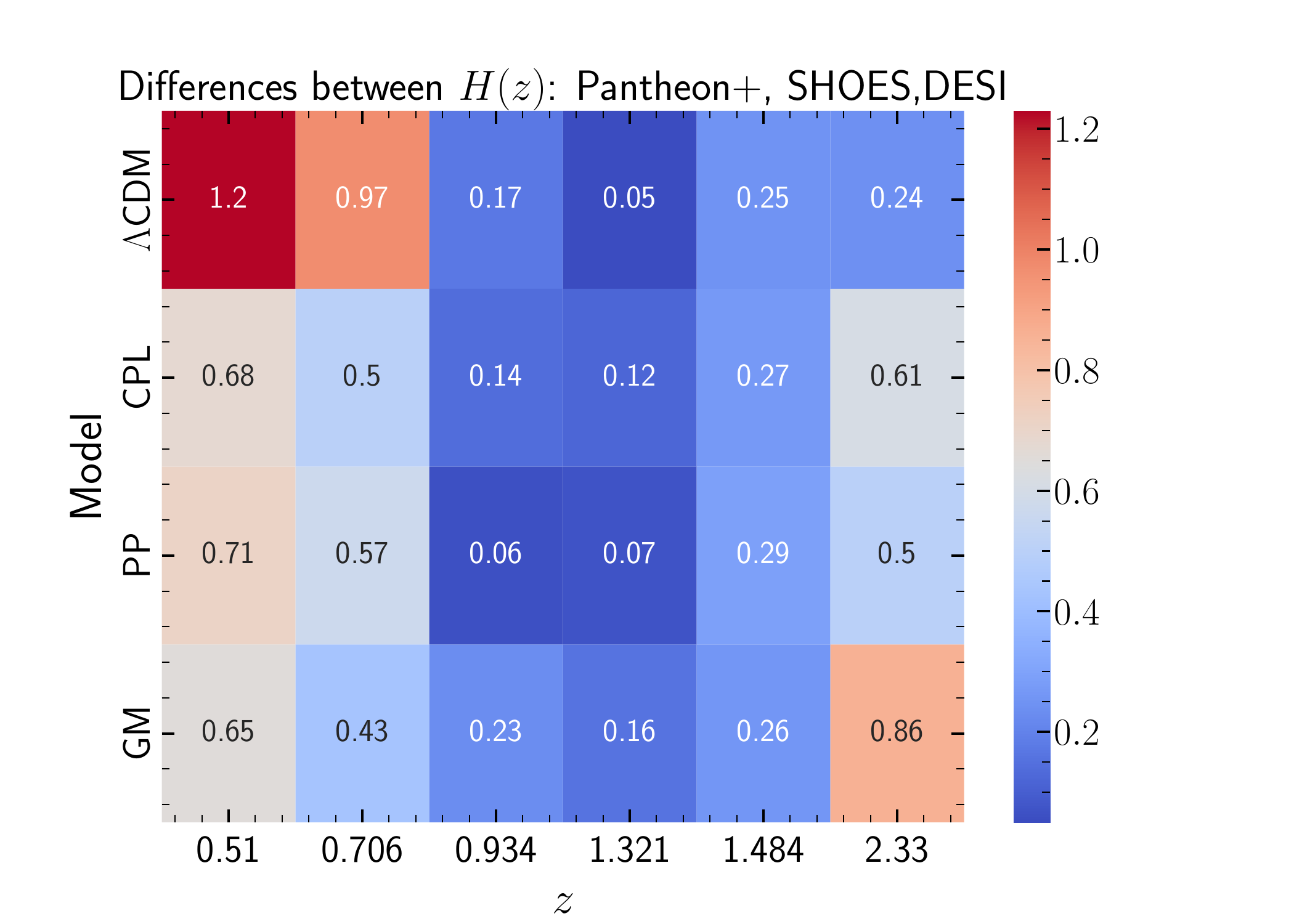}
      \caption{Deviation between GPR reconstructed $H(z)$ and theoretical predictions when Pantheon+SH0ES+DESI is used.}
      \label{fig:heatmap_a}
  \end{subfigure}
  \vspace{0.5cm}
  \begin{subfigure}{0.5\textwidth}
      \centering
      \includegraphics[width=1.0\textwidth]{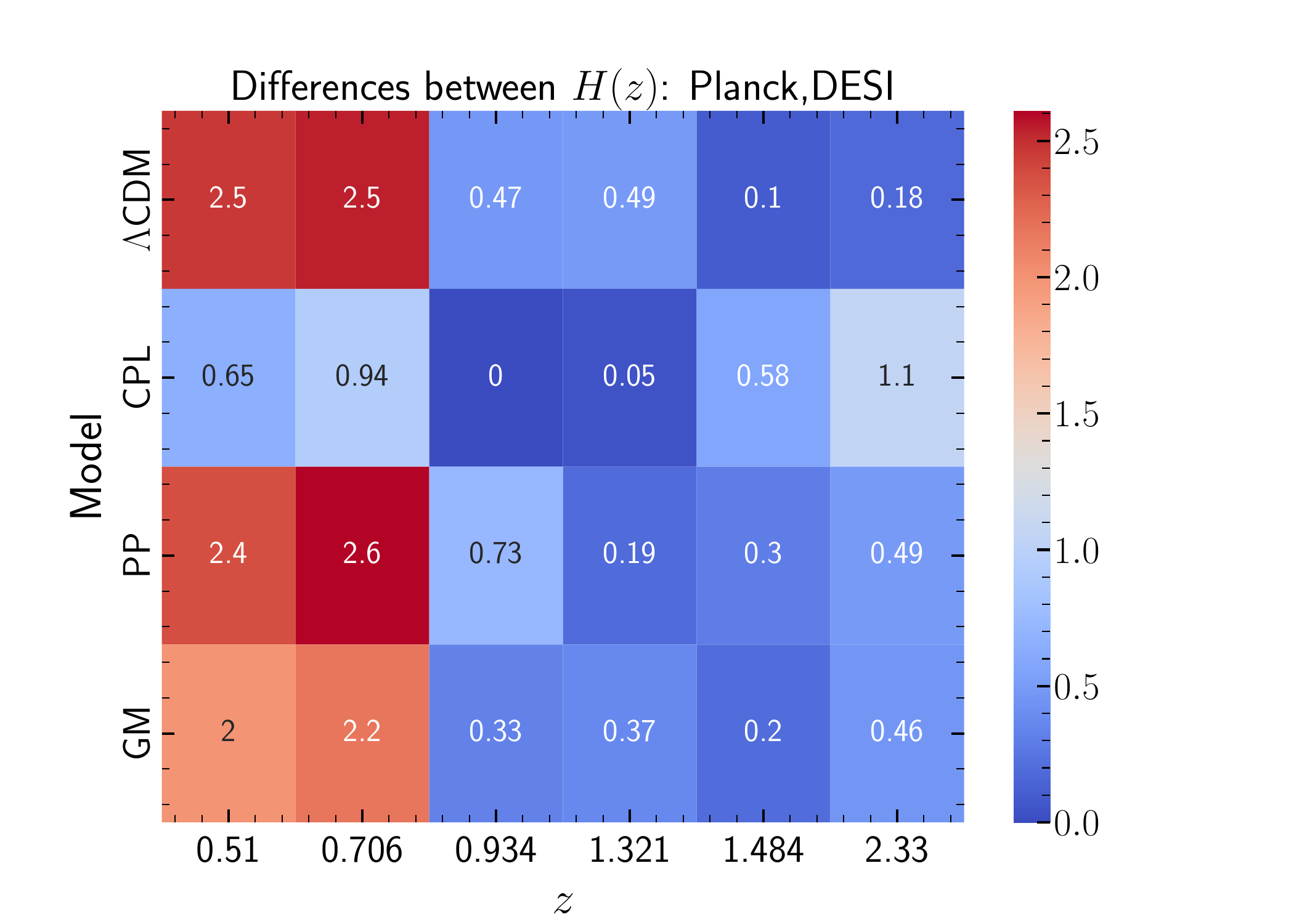}
      \caption{Deviations between GPR reconstructed $H(z)$ and theoretical predictions when Planck+DESI is used.}
      \label{fig:heatmap_b}
  \end{subfigure}
  \caption{Heatmaps showing the statistical significance of deviations between parametric and non-parametric reconstructions of $H(z)$ at redshifts $z = 0.51, 0.706, 0.934, 1.321, 1.484,$ and $2.33$. Predictions are derived within the $\Lambda$CDM, CPL, PP, and GM models. The color scale represents the deviation in units of $\sigma$.}
  \label{heatmap}
\end{figure}

\section{\label{result_DE3} Robustness of the Reconstructed $H(z)$}
The results of Sect.~\ref{result_DE2} show that the discrepancy between the early- and late-time reconstructed expansion histories depends strongly on redshift and on the adopted parametrization. To examine whether this redshift dependence is driven by the assumed functional forms of $\Lambda$CDM, CPL, PP, and GM, we compare their predictions with a non-parametric reconstruction of $H(z)$ obtained using Gaussian Process Regression (GPR). Our aim is not to rank the different cosmological models, but to test whether the redshift localization of the discrepancies identified in Sect.~\ref{result_DE2} persists when $H(z)$ is reconstructed without assuming these parametric forms.

The GPR method is implemented following \cite{Seikel:2012uu}, which has been widely adopted in cosmological studies. Here, the reconstructed function $f(z)$ is modeled as a GP characterized by a mean function $\mu(z)$ and a covariance kernel $k(z,\tilde z)$, which quantifies the correlation between $f(z)$ and $f(\tilde z)$. We perform two independent reconstructions corresponding to the dataset combinations considered in our analysis by adopting the Matérn-(9/2) kernel. For Pantheon+SH0ES+DESI, the expansion history is calibrated using SH0ES, while for Planck+DESI, the DESI measurements are calibrated using the Planck determination of the sound horizon, ($r_d$).

Figure~\ref{heatmap} shows the differences between the parametric predictions of $H(z)$, computed within the $\Lambda$CDM, CPL, PP, and GM frameworks, and the corresponding non-parametric reconstructions at the DESI tracer redshifts $z=0.51$, $0.706$, $0.934$, $1.321$, $1.484$, and $2.33$. 
In Fig.~\ref{heatmap}(a), the non-parametric reconstruction is obtained from Pantheon+SH0ES+DESI, while in Fig.~\ref{heatmap}(b) it is obtained from Planck+DESI. The corresponding parametric predictions are evaluated using Pantheon+SH0ES+DESI and Planck+DESI, respectively.

The heatmaps show that the largest differences generally occur at low redshifts, particularly around $z=0.51$ and $0.706$, while the agreement improves toward higher redshift, where the differences are mostly within $1\sigma$. This behaviour is qualitatively consistent with Sect.~\ref{result_DE2}, where the largest discrepancies for the extended parametrizations were also concentrated at low redshifts. The non-parametric comparison therefore suggests that this localization is not solely a consequence of the assumed functional forms.



To examine this behaviour beyond the discrete DESI tracer redshifts, we also compare the parametric and non-parametric reconstructions continuously over $0\leq z\leq2.5$ and find the resulting differences show the same overall trend, with the largest deviations concentrated toward lower redshift and improved agreement at higher redshift. A similar conclusion is obtained comparing the dimensionless expansion rate, $H/H_0$, further supporting the observed redshift-dependent behaviour.

Some recent analyses have pointed to similar inconsistencies in the DESI DR2 BAO measurements at low redshifts, particularly at $z=0.510$ and $z=0.706$, corresponding to the LRG1 and LRG2 tracer samples \cite{Colgain:2024xqj,Colgain:2025nzf,GuptaChoudhury:2026gsl}.
Interestingly, these redshifts coincide with the region where our parametric vs non-parametric comparison shows the largest differences. This provides an additional consistency check on the redshift dependence identified in Sect.~\ref{result_DE2}.

Moreover, the heatmaps in Fig. \ref{heatmap} reveal that within the considered frameworks the reconstructions anchored to Pantheon+SH0ES+DESI (Fig.~\ref{heatmap} (a)) are systematically in better agreement with the non-parametric reconstructions than those based on Planck+DESI (Fig.~\ref{heatmap} (b)). For instance, at $z=0.706$, the differences for the $\Lambda$CDM, CPL, PP, and GM models in Fig.~\ref{heatmap}(a) are $0.97\sigma$, $0.5\sigma$, $0.57\sigma$, and $0.43\sigma$, respectively, whereas the corresponding differences in Fig.~\ref{heatmap}(b) are $2.5\sigma$, $0.94\sigma$, $2.6\sigma$, and $2.2\sigma$. 

Overall, the non-parametric reconstruction supports the main conclusion of Sect.~\ref{result_DE2}: the discrepancy in the inferred expansion history is not distributed uniformly across redshift, but is most pronounced at low redshifts, particularly around $z=0.510$ and $z=0.706$. The comparison therefore serves as a robustness check on the inferred redshift dependence. 

 \section{\label{Summary} Summary and Discussions}
 The persistent discrepancies between the cosmic expansion rate inferred from independent early- and late-Universe observations have motivated extensive efforts to explain the origin of the Hubble tension. While much attention has been devoted to the discrepancy in the inferred value of $H_0$, identifying the cosmic epoch at which the expansion histories begin to diverge is also important for diagnosing the physical origin of the tension. In this work, we therefore ask whether the expansion histories inferred from early- and late-Universe observations remain discrepant throughout cosmic evolution or whether the mismatch is concentrated within some particular redshift ranges.

Here we adopt generic parametrizations of physical quantities that can govern the expansion history of the Universe. In this regard, we have considered $\Lambda$CDM together with the CPL, PP, and GM parametrizations and have compared the expansion histories inferred from
Pantheon+SH0ES+DESI and Planck+DESI. Further we have compared our predictions with a non-parametric reconstruction of $H(z)$ for a robustness check. The main results of our analysis can be summarized as follows:

 \begin{itemize}
     \item Among the parametric models considered here, the pressure parametrization of DE (PP) emerges as the most promising model for alleviating the $H_0$ tension. It reduces the $H_0$ tension to about $2.6\sigma$ while preserving stable posterior constraints on $\Omega_{m0}$ across the two dataset combinations. The $\Omega_{m0}$ posteriors show substantial overlap, and the corresponding $H_0$--$\Omega_{m0}$ contours show overlap at the boundary of the $2\sigma$ region. 

     \item Within $\Lambda$CDM, the expansion histories reconstructed from early- and late-Universe obsevations remain discrepant at more than 4$\sigma$ over all redshift ranges. In contrast, for CPL, PP, and GM, the dominant differences are concentrated mainly at low redshifts ($z \lesssim 3$), with improved agreement toward higher redshift. 

 \item Although the PP model reduces the $H_0$ tension to approximately $2.6\sigma$, it predicts significant discrepancies between the reconstructed expansion histories at other redshifts. 
 This illustrates that a comparison based only on the present-day expansion rate can therefore miss significant discrepancies that emerge at other redshifts.

 \item The comparison with the non-parametric reconstruction shows a consistent redshift dependence, with the largest differences occurring around $z=0.51$ and $z=0.706$, while the agreement generally improves toward higher redshift. This provides a complementary check on the redshift localization inferred from the parametric reconstructions. 


 \item The reconstructed DE equation of state and energy density in Appendix \ref{appendix A}, also show qualitatively different evolutionary trends for the two dataset combinations. Pantheon+SH0ES+DESI generally favors non-phantom behaviour, whereas Planck+DESI favors phantom behaviour for PP and GM and a phantom crossing for CPL, illustrating the sensitivity of the inferred DE dynamics to the dataset combination \cite{Adi:2025hyj}.
 \end{itemize}

These results show that the mismatch in the reconstructed expansion history from early- and late-time cosmic data is strongly redshift dependent. Within $\Lambda$CDM, the discrepancy in $H(z)$ persists over a broad redshift range, whereas the extended parametrizations localize the dominant differences toward low redshifts.

The results from PP parametrization further illustrates that improved agreement in $H_0$ does not necessarily imply consistency of the full expansion history. A comparison based only on the present-day expansion rate is therefore not a sufficient for establishing the viability of a cosmological model. This broader picture is consistent with recent studies emphasizing the difficulty of fully reconciling current cosmological tensions within homogeneous frameworks and motivates the consideration of additional effects and alternative possibilities \cite{Vagnozzi:2023nrq,Bansal:2026axl,Perivolaropoulos:2025gzo,Yashiki:2025loj,Jedamzik:2025cax,Lee:2025yah,Kamionkowski:2022pkx,Garny:2026ish}.

Both the parametric and non-parametric reconstructions point to the low-redshift regime as the region where the discrepancies are most pronounced. Whether this localization reflects residual calibration or observational systematics, or instead signals a genuine departure from the standard expansion history, remains an open question. Future high-precision measurements from Euclid and the Vera C. Rubin Observatory LSST, together with improved supernova and BAO observations, will therefore be particularly important for determining the origin of these low-redshift discrepancies.

\begin{acknowledgments}
We are grateful to Anjan Ananda Sen, Matteo Gori and Matthieu Sarkis for useful discussion and comments on the draft. AT and UM acknowledge the funding from ERC-AdG Grant ‘FITMOL’ to accomplish the research. PM acknowledges funding from the Anusandhan National Research Foundation (ANRF), Govt of India, under the National Post-Doctoral Fellowship (File no. PDF/2023/001986). PM further acknowledges support from the University of Luxembourg for her visit, where part of this work has been done. This work is based upon discussion from COST Action CA21136 – “Addressing observational tensions in cosmology with systematics and fundamental physics (CosmoVerse)”, supported by COST (European Cooperation in Science and Technology).
\end{acknowledgments}

\appendix
\begin{figure}
\centering
  \begin{subfigure}{0.5\textwidth}  
  \centering
  \includegraphics[width=1.0\textwidth]{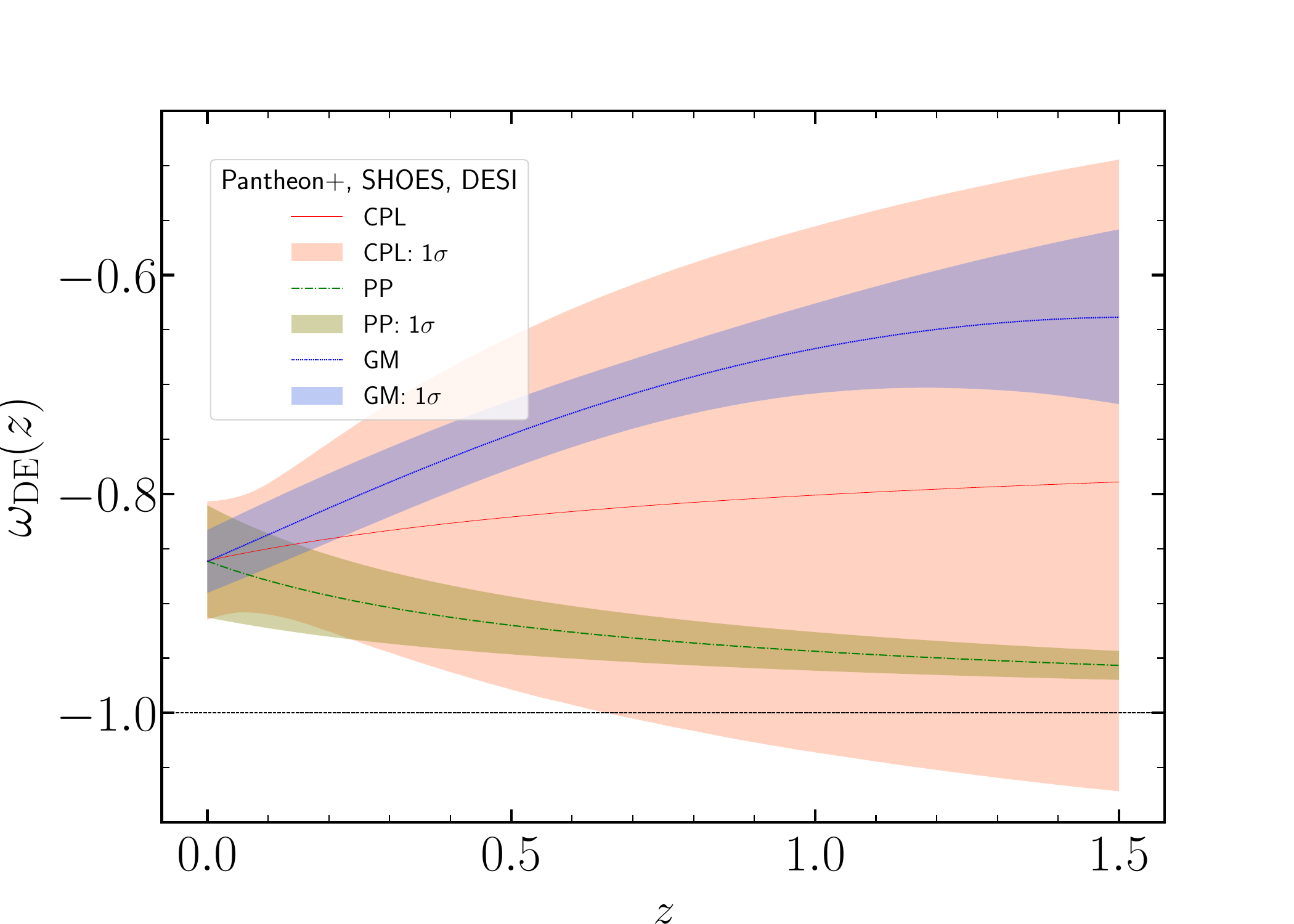}
        \caption{Reconstruction of DE EoS using low-$z$ data combination Pantheon+SH0ES, and DESI.}
  \end{subfigure}
    \vspace{0.5cm}
   \begin{subfigure}{0.5\textwidth}
   \centering
  \includegraphics[width=1.0\textwidth]{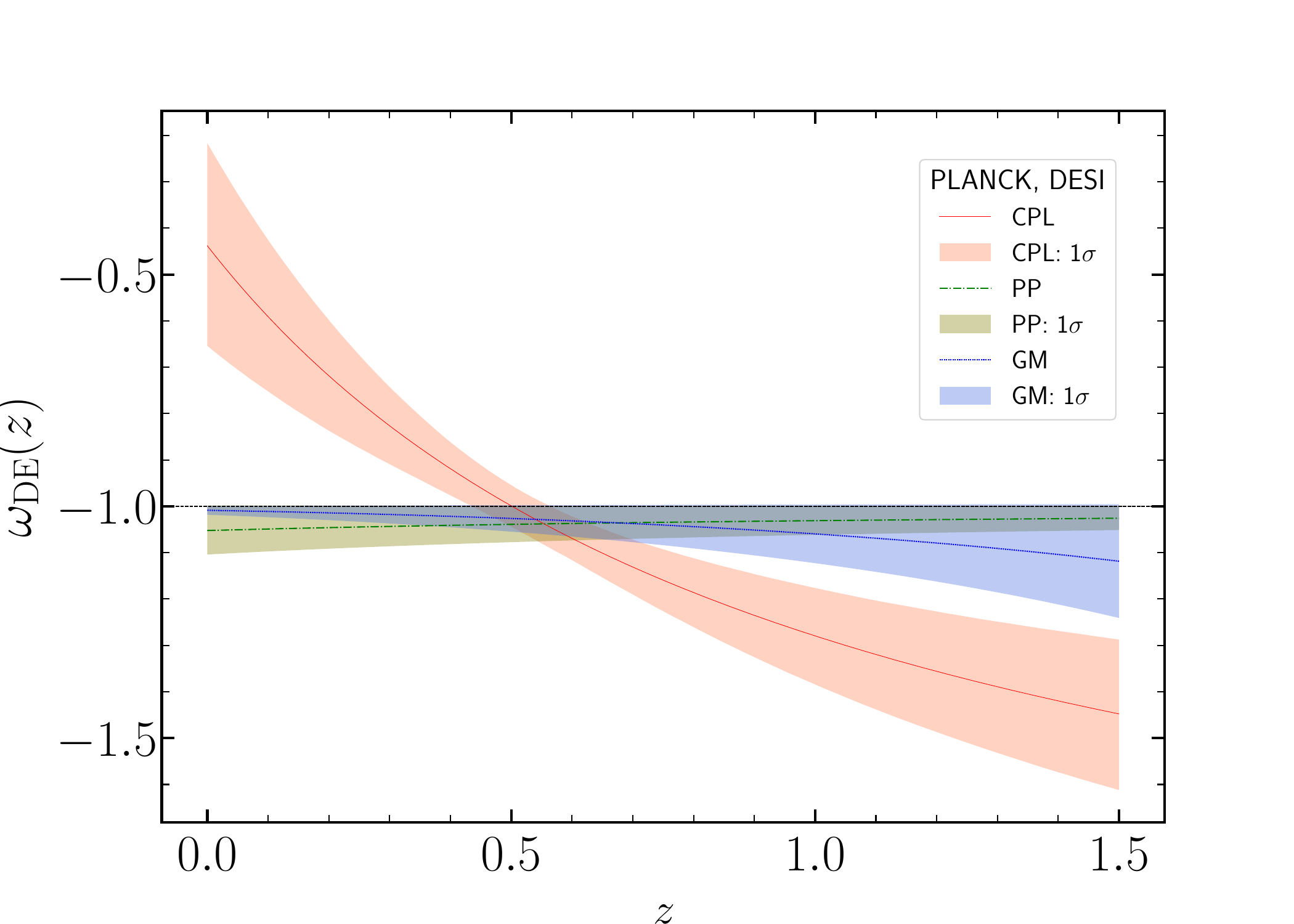}
  \caption{Reconstruction of DE EoS using high-$z$ data combination Planck and DESI.}
    \end{subfigure}
  \caption{Reconstructed DE equation of state $w_{\rm DE}(z)$ for the CPL, PP, and GM models. 
The upper panel shows the reconstruction obtained using low-$z$ data, while the lower panel corresponds to high-$z$ data. 
}
  \label{Fig.EoS}
\end{figure}

\begin{figure}
  \centering
     \begin{subfigure}{0.5\textwidth}
     \centering
  \includegraphics[width=1.0\textwidth]{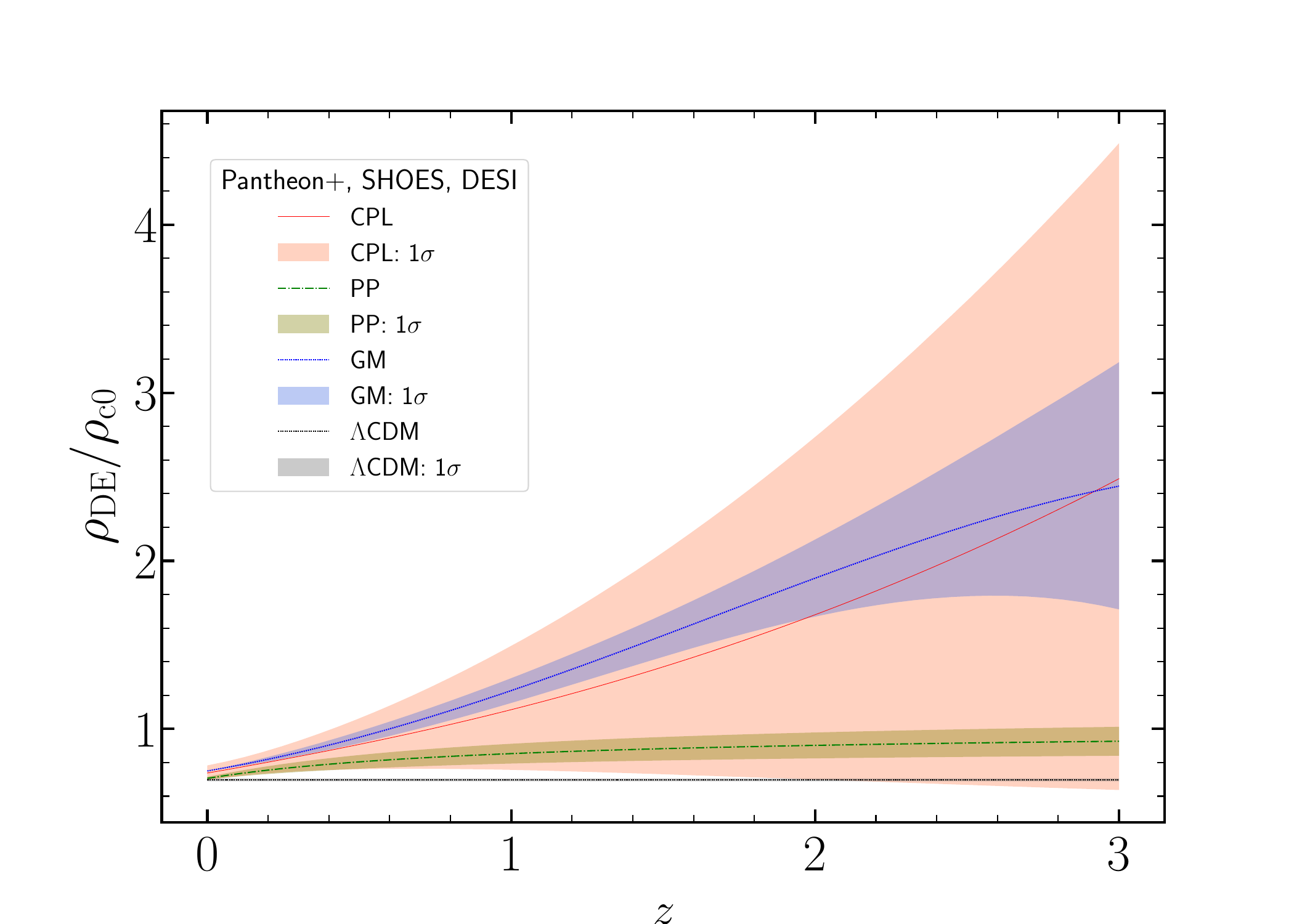}
         \caption{Reconstruction of normalized DE energy density using low-$z$ data combination Pantheon+SH0ES, and DESI.}
  \end{subfigure}
    \vspace{0.5cm}
    \begin{subfigure}{0.5\textwidth}
    \centering
  \includegraphics[width=1.0\textwidth]{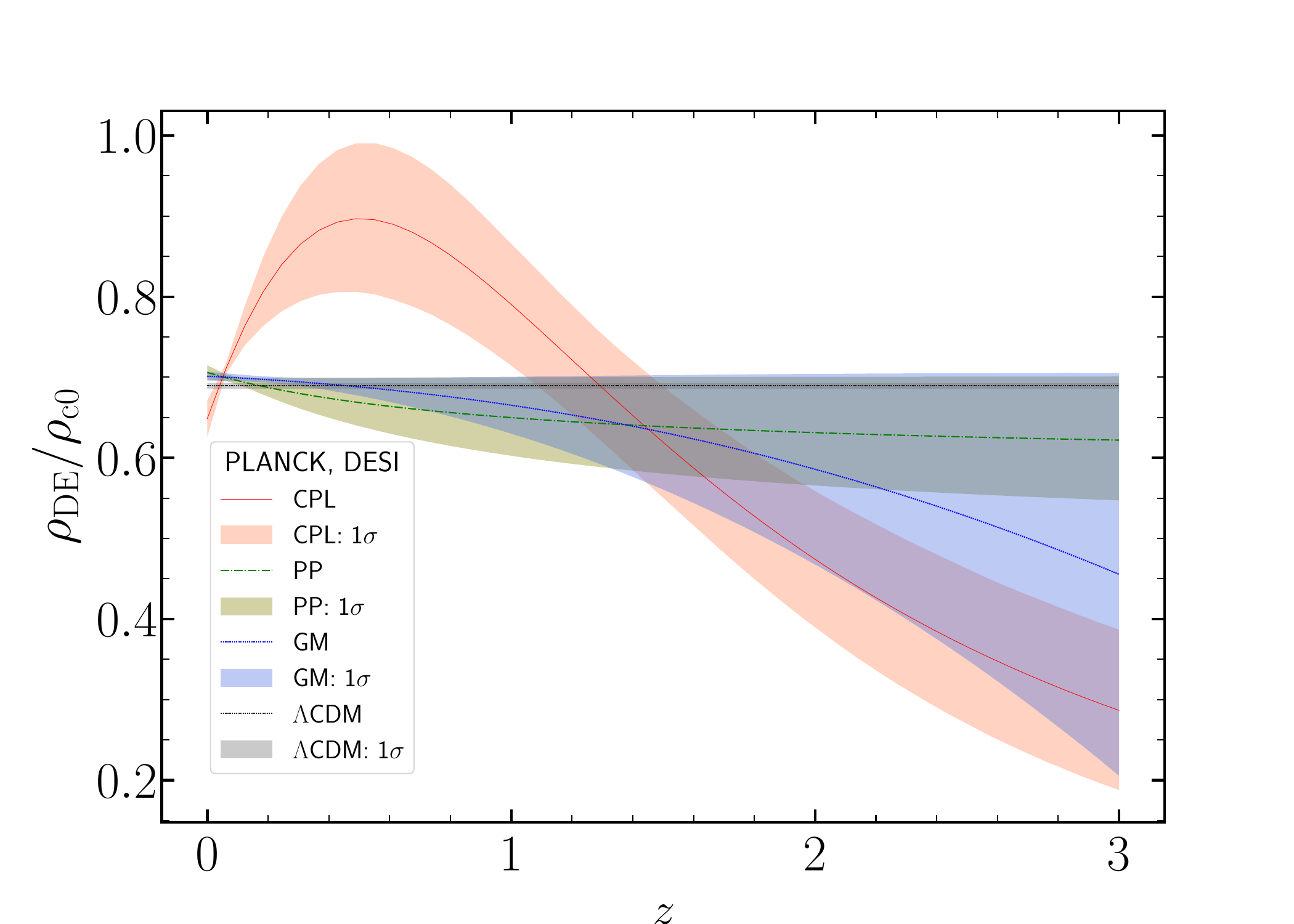}
    \caption{Reconstruction of normalized DE energy density using high-$z$ data combination Planck and DESI.}
      \end{subfigure}
  \caption{Evolution of the DE density normalized to the present-day critical energy density, $\rho_{\rm DE}/\rho_{c0}$, reconstructed for the $\Lambda$CDM, CPL, PP, and GM models. 
The upper panel shows the reconstruction using low-$z$ data, while the lower panel corresponds to high-$z$ data. }
  \label{Fig.OmegaDE}
\end{figure}
\section{Dark Energy Density and Equation of State} \label{appendix A}

In this section, we examine the redshift evolution of the DE equation of state, $w_{\rm DE}$, and energy density, $\rho_{\rm DE}$, for the models introduced in Sect.~\ref{DE_model} and compare the reconstructions obtained from Pantheon+SH0ES+DESI and Planck+DESI.

In Fig.~\ref{Fig.EoS}, we compare the reconstructed DE EoS, $w_{\rm DE}(z)$, for the $\Lambda$CDM, CPL, PP, and GM models. The upper and lower panels correspond to Pantheon+SH0ES+DESI and Planck+DESI, respectively. The red, blue, and green shaded regions represent the best-fit values and $1\sigma$ uncertainties of $w_{\rm DE}$ for the CPL, PP, and GM models, respectively, while $w_{\rm DE}=-1$ denotes the $\Lambda$CDM prediction.
For Pantheon+SH0ES+DESI, PP exhibits a thawing behaviour~\cite{Caldwell:2005tm}, approaching $w_{\rm DE}=-1$ at higher redshift and deviating from it toward lower redshift. Among the dynamical models, PP remains closest to $\Lambda$CDM over the reconstructed redshift range, but its departure from $w_{\rm DE}=-1$ reflects the dynamical nature of DE. The GM model shows a larger departure from $w_{\rm DE}=-1$ at higher redshift and evolves toward values closer to $-1$ at lower redshift. For the GM parametrization, $\Omega_{m0}$ cannot be robustly constrained using Pantheon+SH0ES+DESI alone. We therefore adopt its value from Ref.~\cite{Mukhopadhyay:2024fch}. 
Varying $\Omega_{m0}$ within reasonable bounds shifts $w_{\rm DE}(z)$ vertically without qualitatively changing its redshift evolution. PP and GM also yield tighter $w_{\rm DE}$ constraints than CPL.

For Planck+DESI, CPL exhibits a phantom crossing near $z\sim0.5$, transitioning from $w_{\rm DE}<-1$ to $w_{\rm DE}>-1$. In contrast, PP and GM remain in the phantom regime over the reconstructed redshift range. PP remains nearly constant, whereas GM departs progressively from $w_{\rm DE}=-1$ toward higher redshift. The PP model can also exhibit phantom crossing when extended to second order \cite{Cheng:2025lod}.
These results show that the phantom-crossing behaviour reported in recent DESI analyses~\cite{DESI:2025fii} is sensitive to the adopted parametrization and is therefore not generic to all phenomenological descriptions of dynamical DE (see Refs.~\cite{Wolf:2024eph,Gialamas:2025pwv,Gialamas:2024lyw,Shlivko:2024llw}).

Figure~\ref{Fig.OmegaDE} shows the corresponding evolution of $\rho_{\rm DE}/\rho_{c0}$. 
For Pantheon+SH0ES+DESI, CPL and GM show a decrease in $\rho_{\rm DE}/\rho_{c0}$ toward lower redshift, while PP remains nearly constant. 
This behaviour is consistent with the predominantly non-phantom evolution inferred from Fig.~\ref{Fig.EoS}. For Planck+DESI, $\rho_{\rm DE}/\rho_{c0}$ increases toward lower redshift for PP and GM, consistent with their phantom behaviour. CPL follows a different evolution associated with its phantom crossing, transitioning from $w_{\rm DE} < -1$ to $w_{\rm DE} > -1$ at $z \sim 0.5$. 

Taken together, Figs.~\ref{Fig.EoS} and \ref{Fig.OmegaDE} show that the inferred DE evolution depends on both the adopted parametrization and the dataset used to constrain it. Pantheon+SH0ES+DESI generally favors non-phantom evolution, whereas Planck+DESI favors phantom behaviour for PP and GM and a phantom crossing for CPL. This contrast highlights a mismatch between the two dataset combinations in characterizing the dynamics of DE.

\bibliography{aapsamp}

\end{document}